\documentclass[fleqn,12pt,twoside]{article}
\usepackage{espcrc1}

\usepackage{graphicx}

\makeatletter
\def\simleq{\mathrel{\mathpalette\gl@align<}}
\def\simgeq{\mathrel{\mathpalette\gl@align>}}
\def\gl@align#1#2{\lower.6ex\vbox{\baselineskip\z@skip\lineskip\z@
     \ialign{$\m@th#1\hfill##\hfil$\crcr#2\crcr\sim\crcr}}}
\makeatother

\newcommand{\AmS}{{\protect\the\textfont2
  A\kern-.1667em\lower.5ex\hbox{M}\kern-.125emS}}

\title{
Gluonic Higgs Scalar, Abelianization and Monopoles in QCD
-- Similarity and Difference between QCD in the MA Gauge and the NAH Theory
}

\author{Hideo Suganuma\address[titech]{
			Tokyo Institute of Technology,
			Oh-Okayama 2-12-1, Meguro, Tokyo 152-8551, Japan} and
	Hiroko Ichie\address[humboldt]{Humboldt Univ. zu Berlin, 
Inst. f\"ur Phys., Invalidenstrasse, D-10115 Berlin, Germany}
}
       
\begin{document}

\maketitle

\begin{abstract}
We study the similarity and the difference between QCD in the 
maximally abelian (MA) gauge and the nonabelian Higgs (NAH) theory 
by introducing the ``gluonic Higgs scalar field'' 
$\vec \phi(x)$ corresponding to the ``color-direction'' of the nonabelian gauge 
connection. 
The infrared-relevant gluonic mode in QCD can be extracted by the projection 
along the color-direction $\vec \phi(x)$ like the NAH theory.
This projection is manifestly gauge-invariant, and 
is mathematically equivalent to the ordinary MA projection.
Since $\vec \phi(x)$ obeys the adjoint gauge transformation  
and is diagonalized in the MA gauge, 
$\vec \phi(x)$ behaves as the Higgs scalar in the NAH theory, and 
its hedgehog singularity provides the magnetic monopole in the MA gauge 
like the NAH theory. 
We observe this direct correspondence between the monopole appearing in the MA gauge 
and the hedgehog singularity of $\vec \phi(x)$ in lattice QCD, 
when the gluon field is continuous as in the SU($N_c$) Landau gauge.
In spite of several similarities, QCD in the MA gauge largely differs from the NAH theory in the two points: 
one is infrared monopole condensation, and the other is infrared enhancement of the abelian correlation
due to monopole condensation.
\end{abstract}

\section{QCD in the MA Gauge and Dual Superconductor Theory for Confinement}

To understand the confinement mechanism is 
one of the most difficult problems remaining in the particle physics~\cite{conf2000}.
Quark confinement is 
characterized by one-dimensional squeezing of the 
color-electric flux with  
the string tension $\sigma \simeq 1{\rm GeV/fm}$, 
which is the universal key quantity in QCD~\cite{TS}.
On the confinement mechanism, based on the electro-magnetic duality, 
Nambu~\cite{N} first proposed the dual superconductor theory,  
where the one-dimensional squeezing of the color-electric flux  
occurs by the dual Meissner effect due to condensation of 
bosonic color-magnetic monopoles. 
But, there are {\it two large gaps} between QCD and the 
dual superconductor theory.
\begin{enumerate}
\item
The dual superconductor theory is based on the {\it abelian gauge theory} 
subject to the Maxwell-type equations, where electro-magnetic duality is 
manifest, while QCD is a nonabelian gauge theory.   
\item
The dual superconductor theory requires {\it color-magnetic monopole condensation} 
as the key concept, while QCD does not have 
color-magnetic monopoles as the elementary degrees of freedom.
\end{enumerate}
These gaps may be filled simultaneously by taking 
maximally abelian (MA) gauge fixing, which reduces QCD to an abelian gauge theory 
including color-magnetic monopoles.

In Euclidean QCD, the MA gauge is 
defined so as to minimize the ``total amount" of the 
off-diagonal gluon amplitude, 
\begin{equation}
R_{\rm off} [A_\mu ( \cdot )] \equiv \int d^4x \ {\rm tr}
\left\{ 
[\hat D_\mu ,\vec H][\hat D_\mu ,\vec H]^\dagger 
\right\} 
\end{equation}
by the SU($N_c$) gauge transformation~\cite{AS,IS,S,SI}.
In the MA gauge, the nonabelian gauge symmetry is partially fixed as 
$G \equiv {\rm SU}(N_c)_{\rm local} \rightarrow  
H \equiv {\rm U(1)}_{\rm local}^{N_c-1} 
\times {\rm Weyl}^{\rm global}_{N_c}$, 
and QCD reduces into an abelian gauge theory. 
Furthermore, according to the reduction of the gauge symmetry, 
color-magnetic monopoles appear as the topological defects reflecting 
the nontrivial homotopy group 
$
\Pi_2({\rm SU}(N_c)/{\rm U(1)}^{N_c-1})=\Pi_1({\rm U(1)}^{N_c-1})
={\bf Z}^{N_c-1}_\infty 
$
~\cite{IS,S,SI,SST,thooft} 
in a similar manner to the appearance of the 't~Hooft-Polyakov monopole 
in the NAH theory.

As remarkable features, lattice QCD in the MA gauge exhibits   
{\it infrared abelian dominance} and {\it infrared monopole condensation}~\cite{AS,IS,S,SI}, 
which provide a theoretical basis of the 
dual superconductor theory for quark confinement.
 
\section{The similarity between QCD in the MA gauge and the Nonabelian Higgs theory: 
Appearance of Magnetic Monopoles and Infrared Abelianization}

\begin{figure}
\begin{center}
\includegraphics[scale=0.3]{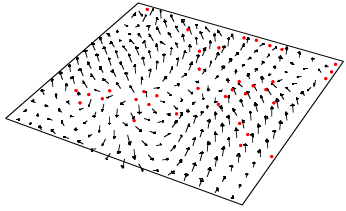}
\vspace{-0.5cm}
\caption{\label{hedgehog}
The gluonic Higgs scalar field $\phi(x)=\phi^a(x)\frac{\tau^a}{2}$ 
in the SU(2) Landau gauge in SU(2) lattice QCD 
with $\beta=2.4$ and $16^4$.
The arrow denotes the SU(2) color direction,  
$(\phi^1(x),\phi^2(x),\phi^3(x))$.
The monopoles (dots) in the MA gauge appear at the hedgehog singularities of 
the gluonic Higgs scalar $\phi(x)$.}
\end{center}
\vspace{-0.5cm}
\end{figure}

QCD in the MA gauge is similar to the NAH theory in terms of 
appearance of magnetic monopoles. 
To clarify the similarity between QCD in the MA gauge and the NAH theory, 
we introduce the ``gluonic Higgs scalar field'' $\vec \phi(x)$ \cite{IS,S}
as a function of the gluon-field configuration $\{A_\mu(x)\}$. 
For arbitrary given gluon configuration $\{ A_\mu(x) \}$, 
we define 
$\vec \phi(x) \equiv \Omega(x) \vec H \Omega^\dagger(x)$ with
$\Omega(x) \in {\rm SU}(N_c)$ so as to minimize 
\begin{eqnarray}
R[\vec \phi(\cdot)] \equiv \int d^4x \ {\rm tr} 
\left\{[\hat D_\mu, \vec \phi(x)][\hat D_\mu, \vec \phi(x)]^\dagger \right\}
\end{eqnarray}
in the Euclidean metric. 
The gluonic Higgs scalar 
$\vec \phi(x)$ physically corresponds to 
the ``color-direction'' of the nonabelian gauge connection $\hat D_\mu$ 
averaged over $\mu$ at each $x$. 

Similar to the covariant derivative $\hat D_\mu$, the gluonic Higgs scalar 
$\vec \phi(x)$ obeys the {\it adjoint gauge transformation} as $\vec \phi(x) \rightarrow V^\dagger(x) \vec\phi(x) V(x)$, 
and is diagonalized in the MA gauge. 
Therefore, $\vec \phi(x)$ behaves as the Higgs scalar in the NAH theory, and 
the hedgehog singularity of $\vec \phi(x)$ provides the magnetic monopole in the MA gauge~\cite{IS,S,SI,SST}, 
as a direct analogue of the appearance of the 't~Hooft-Polyakov monopole in the SU($N$) NAH theory.  

Actually in lattice QCD, we observe this direct correspondence between the monopole appearing in the MA gauge 
and the hedgehog singularity of $\vec \phi(x)$ \cite{IS,S,SI,SST}, 
when the gluon field is continuous as in the SU($N_c$) Landau gauge, as shown in Fig.~1.
(In the SU($N_c$) Landau gauge, the gauge field is maximally continuous, 
so that the correspondence between the hedgehog singularity and the monopole position become manifest. 
However, in the original random gauge on lattice, such correspondence cannot be observed.)

Note here that the adjoint gauge-transformation property of $\vec \phi(x)$ 
is essential on the correspondence between the hedgehog singularity of $\vec \phi(x)$ 
and the monopole singularity appearing in the MA gauge. 
For instance, consider the other operator composed by the link-variable $U_\mu(s)$,  
\begin{eqnarray}
X(s) \equiv 
\sum_{\mu=1}^4 \{U_{\mu}(s)\tau_3U_{\mu}^\dagger(s)
+U_\mu^\dagger(s-\hat\mu)\tau_3U_\mu(s-\hat\mu)\}, 
\end{eqnarray} 
which corresponds to $X(x) =[\hat D_\mu, [\hat D_\mu, \tau_3]]$ in the continuum limit.
In SU(2) QCD, $X(s)$ is also diagonalized in the MA gauge, but it does not obey the adjoint transformation 
by the gauge transformation, so that there is no simple correspondence between the hedgehog singularity 
of $X(s)$ and the monopole position in the MA gauge. 

\begin{figure}
\begin{center}
\includegraphics[scale=0.3]{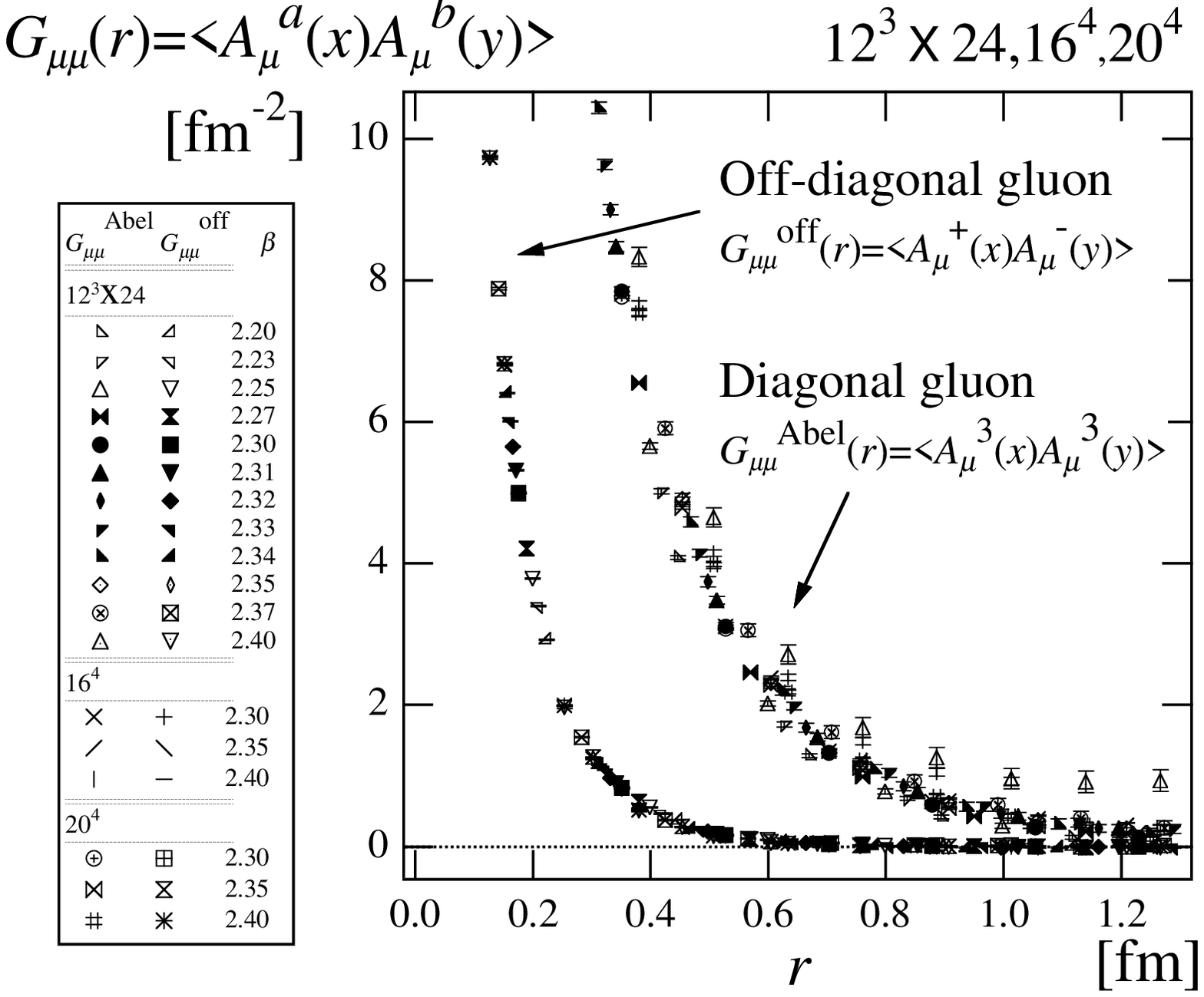}
\includegraphics[scale=0.3]{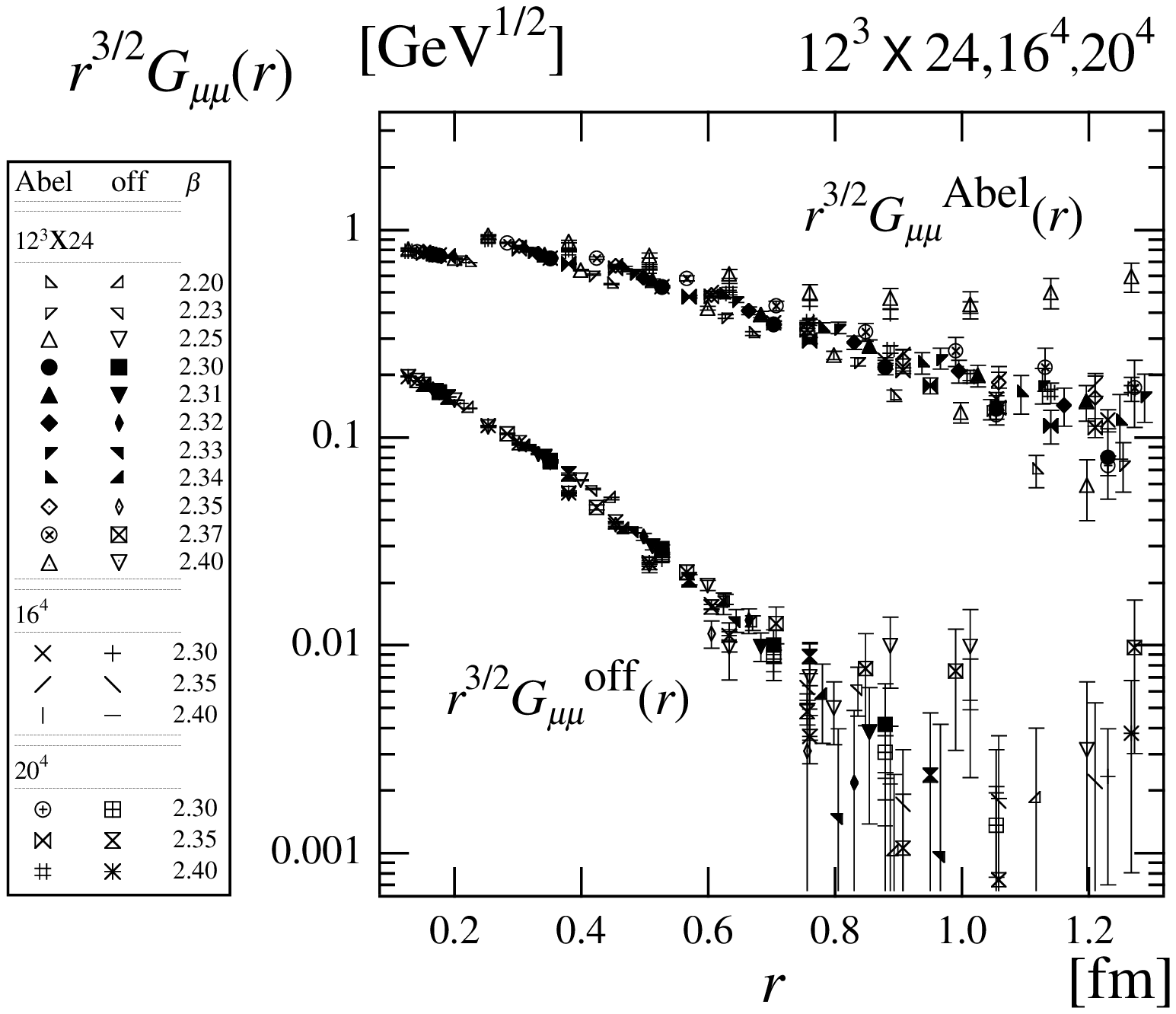}
\includegraphics[scale=0.7]{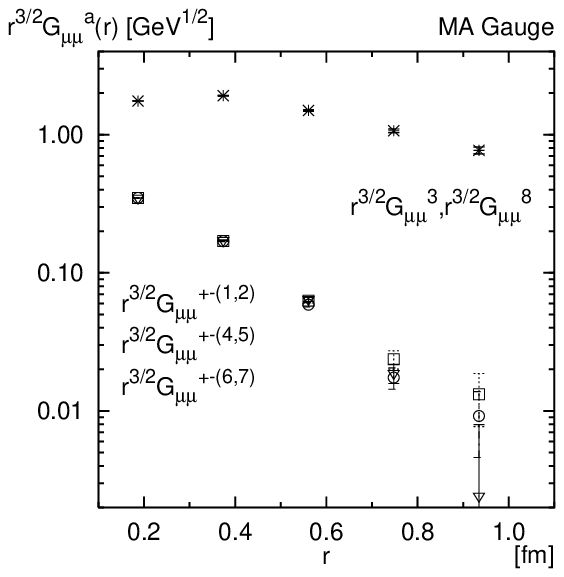}
\vspace{-0.65cm}
\caption{\label{gluonprop}
(a) The scalar-type Euclidean gluon propagator 
$G_{\mu \mu }^a(r)\equiv\langle A_\mu^a(x)A_\mu^a(y) \rangle$ plotted against four-dimensional distance $r \equiv \{(x-y)^2\}^{1/2}$ 
in the MA gauge in SU(2) lattice QCD. 
(b) The logarithmic plot of $r^{3/2} G_{\mu \mu}^a(r)$ in SU(2) lattice QCD. 
(c) The logarithmic plot of $r^{3/2} G_{\mu \mu}^a(r)$ on   
the gluon propagator 
$G_{\mu \nu }^a(r)$ in the MA gauge in SU(3) lattice QCD. 
}
\end{center}
\vspace{-0.6cm}
\end{figure}

Next, let us consider the massive behavior of  off-diagonal gluons and 
resulting infrared abelianization in QCD in the MA gauge.  
Like the off-diagonal (charged) gauge fields in the NAH theory, 
the off-diagonal gluons behave as massive vector fields with a large mass of 
about 1GeV in QCD in the MA gauge with the abelian Landau gauge \cite{AS,S,SI}, as shown in Fig.~2.
Therefore, through the projection along $\vec \phi(x)$, 
one can extract the abelian U(1)$^{N_c-1}$ sub-gauge-manifold 
which is close to the original SU($N_c$) gauge manifold. 
This projection is manifestly gauge-invariant and is 
mathematically equivalent to the ordinary MA projection. 
In fact, the {\it infrared-relevant gluonic mode in QCD can be extracted by the projection 
along the color-direction $\vec \phi(x)$ like the NAH theory} \cite{SI}, 
with the similar argument to infrared abelian dominance in the MA gauge.

Thus, we have the similarities between QCD in the MA gauge and the NAH theory on 
the appearance of magnetic monopoles, the massive behavior of   
off-diagonal gluons and infrared abelianization. 

\section
{Difference between QCD in the MA gauge and the NAH theory: 
Infrared Monopole Condensation and Infrared Enhancement 
of Abelian Correlation}

So far, we have shown the similarities between QCD in the MA gauge and the NAH theory. 
Of course, these theories are essentially different  on the points how the gauge symmetry is broken: 
\begin{enumerate}
\item
While the spontaneous gauge-symmetry breaking occurs in the NAH theory, 
the MA gauge is brought as a gauge fixing in QCD.
\item
While the NAH theory has the Higgs scalar as an elementary field, 
the gluonic Higgs scalar $\vec \phi(x)$ is a composite field of gluons.
\end{enumerate}
Except for these trivial differences, QCD in the MA gauge largely differs from the NAH theory on the following two points:
one is {\it infrared monopole condensation,} and the other is {\it infrared enhancement of the abelian correlation} 
due to monopole condensation.
\begin{enumerate}
\item
Infrared monopole condensation occurs in QCD in the MA gauge, while the magnetic monopole appears as  
an ordinary massive particle in the NAH theory.
\item
Infrared enhancement of abelian correlation is caused by monopole condensation, and provides 
the linear potential at large distances, which leads to the quark confinement, 
while the NAH theory only provides the Coulomb potential at large distances.
\end{enumerate}
We conjecture that infrared monopole condensation occurs as a result of 
the large quantum fluctuation of gluon fields in the infrared region, reflecting the asymptotic freedom.
In fact, the gluonic Higgs field $\vec \phi(x)$ corresponding to 
the color-direction of the nonabelian gauge connection $\hat D_\mu$ is expected to be largely fluctuated in QCD at a large scale, 
and this fluctuation would lead to stochastic monopole excitations 
which may  be interpreted as infrared monopole condensation.

\end{document}